\documentclass[12pt]{article}
\usepackage[T2A]{fontenc}
\usepackage[utf8]{inputenc}
\usepackage[english]{babel}
\usepackage{amssymb}
\usepackage{amsmath}
\usepackage{graphicx}
\usepackage[small,sc,center]{titlesec}
\usepackage{indentfirst}
\usepackage{siunitx}
\usepackage[labelsep=period]{caption}
\usepackage{caption}

\newcommand{\nc}{\newcommand}
\nc{\ek}{E_\mathrm{K}}
\nc{\menv}{M_\mathrm{env}}
\nc{\teff}{T_\mathrm{eff}}
\nc{\tev}{t_\mathrm{ev}}

\begin{document}

\begin{center}
\textbf{Evolution and pulsations of population I post--AGB stars}

\vskip 3mm
\copyright\quad
\textbf{2019 г. \quad Yu. A. Fadeyev\footnote{E--mail: fadeyev@inasan.ru}}

\textit{Institute of Astronomy, Russian Academy of Sciences,
        Pyatnitskaya ul. 48, Moscow, 119017 Russia} \\

Received June 4, 2019; revised June 11, 2019; accepted June 25, 2019
\end{center}

\textbf{Abstract} ---
Evolutionary calculations of population I stars with initial masses $M_0=1M_\odot$, $1.5M_\odot$
and $2M_\odot$ were carried out up to the stage of the proto--planetary nebula.
Selected models of post--AGB evolutionary sequences with effective temperatures
$3.6\times 10^3\,\mathrm{K}\lesssim\teff\lesssim 2\times 10^4\,\mathrm{K}$
were used as initial conditions in calculations of self--escited stellar oscillations.
For the first time the sequences of hydrodynamic models of radially pulsating post--AGB stars
were computed using the self--consistent solution of the equations of radiation hydrodynamics
and time--dependent convection.
Within this range of effective temperatures the post--AGB stars are the fundamental mode
pulsators with period decreasing as the star evolves from $\Pi\approx 300$ day to several days.
Period fluctuations are due to nonlinear effects and are most prominent at effective
temperatures $\teff < 5000$\,K.
The amplitude of bolometric light variations is $\Delta M_\mathrm{bol}\approx 1$ at
$\teff \lesssim 6000$\,K and rapidly decreases with increasing $\teff$.
The theoretical dependence of the pulsation period as a function of effective temperature
obtained in the study can be used as a criterion for the evolutionary status of pulsating
variables suspected to be post--AGB stars.


\newpage
\section*{introduction}

The red giant evolutionary stage of stars with solar composition and the zero--age main sequence
mass $M_0 \lesssim 9M_\odot$ is completed due to the strong stellar wind and the loss of the major
fraction of the hydrogen envelope.
The star leaves the asymptotic giant branch (AGB) and crosses the Hertzsprung--Russel
diagram (HRD) at nearly constant luminosity towards the region of planetary nebula cores
with effective temperatures $\teff\sim 10^5$\, K.
An idea that the planetary nebulae originate from red giants was suggested by Shklovsky (1956)
and was later confirmed by evolutionary computations
(Paczy\'nski, 1971; Sch\"onberner, 1979; 1981; Wood, Faulkner, 1986; Vassiliadis, Wood, 1994;
Bl\"ocker, 1995; Weiss, Ferguson, 2009; Miller Bertolami, 2016).

Formation of the opaque gas--dust envelope surrounding the star on the tip of the AGB
substantially restricts abilities of optical observations, so that the evolutionary
transition to the post--AGB stage still remains unclear.
The strong stellar wind on the tip of the AGB seems to be due to nonlinear stellar pulsations
that lead to dynamical instability of the outer stellar layers (Tuchman et al., 1978).
Therefore, large infrared (IR) excesses detected in post--AGB stars
(Volk, Kwok, 1989; Hrivnak et al., 1989; 1994; Ikonnikova et al., 2018)
indicate existence of circumstellar dust grains condensed during the preceding evolutionary
stage with high mass loss rates.

The photometric variability due to stellar pulsations is a typical property of post--AGB stars.
The period of light variations ranges from several dozen days
(Arkhipova et al., 2010; Hrivnak et al.2015; 2018)
to $\Pi\approx 200$ day (Arkhipova et al., 2009; 2016; Ikonnikova et al., 2018).
Moreover, the pulsating variable AI~CMi with period $\Pi\approx 310$ day seems to be
the early post--AGB star (Arkhipova et al., 2017).
A common feature of post--AGB stars is a lack of strict repetition in light variations,
so that observational estimates of the period are often highly uncertain.

The post--AGB stars are on the late stage of stellar evolution so that comparison of observations
with results of evolutionary computations is often accompanied with difficulties arising from
uncertainties of the theory (e.g., the mass loss rate on the AGB stage).
This problem can be clarified with application of the stellar pulsation theory.
Results of the linear analysis of pulsational instability were presented by
Zalewski (1985; 1993) and Gautschy (1993) who showed that the region of pulsational instability
of post--AGB stars on the HRD extends to effective temperatures as high as $\teff\approx 10^4$\, K.
However one should bear in mind that at luminosity to mass ratios typical for post--AGB stars
($L/M\sim 10^4L_\odot/M_\odot$) the low gas density and the small adiabatic exponent in the
pulsating envelope imply high nonadiabaticity and strong nonlinearity of stellar oscillations.
Results of hydrodynamic computations show that radial oscillations of low--mass supergiants
of intermediate spectral types have the large amplitude and absence of strict periodicity
is due to nonlinear effects (Fadeyev, Tutukov, 1981; Fadeyev, 1982; 1984;
Aikawa, 1985a; 1985b; 1991; 1993; Fokin et al., 2001).
Unfortunately, in all these works the authors ignored convection and assumed that energy transfer
in the pulsating envelope is due to radiation.
However the role of convection in stellar pulsation motions becomes important
at effective temperatures $\teff < 5000$\,K, i.e. at pulsation periods $\Pi \gtrsim 50$~day.

The goal of the present work is to investigate nonlinear oscillations of post--AGB stars with
effective temperatures $3.6\times 10^3\,\mathrm{K}\le\teff\le 2\times 10^4\,\mathrm{K}$.
The equations of radiation hydrodynamics and time--dependent convection are solved
with initial conditions obtained from selected models of previously calculated evolutionary
sequences.
Computations of stellar evolution were done with the MESA code version 10398
(Paxton et al., 2018) from the main sequence to the stage of the proto--planetary nebula with
effective temperature $\teff\approx 2\times 10^4$\, K.
Details of evolutionary computations (the nuclear network, convective mixing, mass loss rates)
are described in our previous study (Fadeyev, 2018).
The initial composition of stellar material was assumed to correspond to population~I stars
with fractional mass abundances of hydrogen and helium $X=0.70$ and $Y=0.28$, respectively.
Abundances of elements heavier than helium were scaled according to 
Grevesse and Sauval (1998).
Equations of hydrodynamics and parameters of the time--dependent convection theory (Kuhfu\ss, 1986)
are discussed in our earlier papers (Fadeyev, 2013; 2015).

\section*{evolutionary sequences and initial conditions}

\begin{figure}
\centerline{\includegraphics[width=14cm]{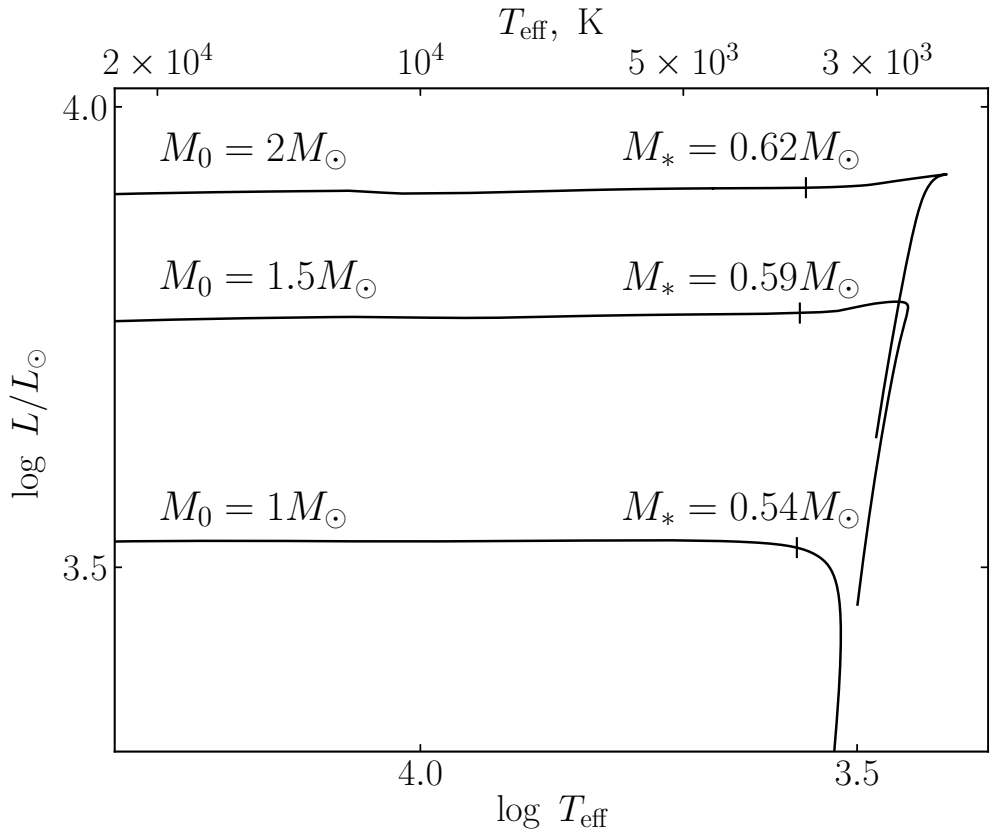}}
\caption{Evolutionary tracks of stars with initial masses $M_0=1M_\odot$, $1.5M_\odot$ and $2M_\odot$
         during transition from AGB to post--AGB.
         $M_*$ is the stellar mass at the onset of the post--AGB stage indicated on the plots
         by vertical dashes.}
\label{fig1}
\end{figure}

In the present study we computed three evolutionary sequences of stars with initial masses
on the main sequence $M_0=1M_\odot$, $1.5M_\odot$ and $2M_\odot$.
Evolutionary tracks used in the present study for determination of initial conditions
are shown in Fig.~\ref{fig1}.
Increase of luminosity in the right part of the figure represents the final phase of the AGB
stage after the last thermal flash of the helium burning shell source.
During this phase the mass of the hydrogen envelope $\menv$ rapidly decreases, whereas
the stellar luminosity is generated in the hydrogen burning shell.
According to Miller Bertolami (2016) we assume that the the post--AGB stage begins
when the hydrogen envelope mass to star mass ratio is $\menv/M = 0.01$.
The onset of the post--AGB stage is indicated for each track in Fig.~\ref{fig1} by the vertical dash
where for the sake on convenience the evolutionary time is set to zero ($\tev=0$).
Further stellar evolution proceeds towards higher effective temperatures at insignificantly
changing luminosity.

Main properties of stars at the beginning of the post--AGB stage are listed in Table~\ref{tabl1},
where $M_0$ is the initial stellar mass, $t_*$ is the star age measured from the zero age
main sequence, $M_*$ is the star mass, $M_{\mathrm{CO},*}$ is the mass of the degenerate
carbon--oxygen core, $L_*$ is the stellar luminosity, $T_{\mathrm{eff},*}$ is the
effective temperature.

\begin{figure}
\centerline{\includegraphics[width=14cm]{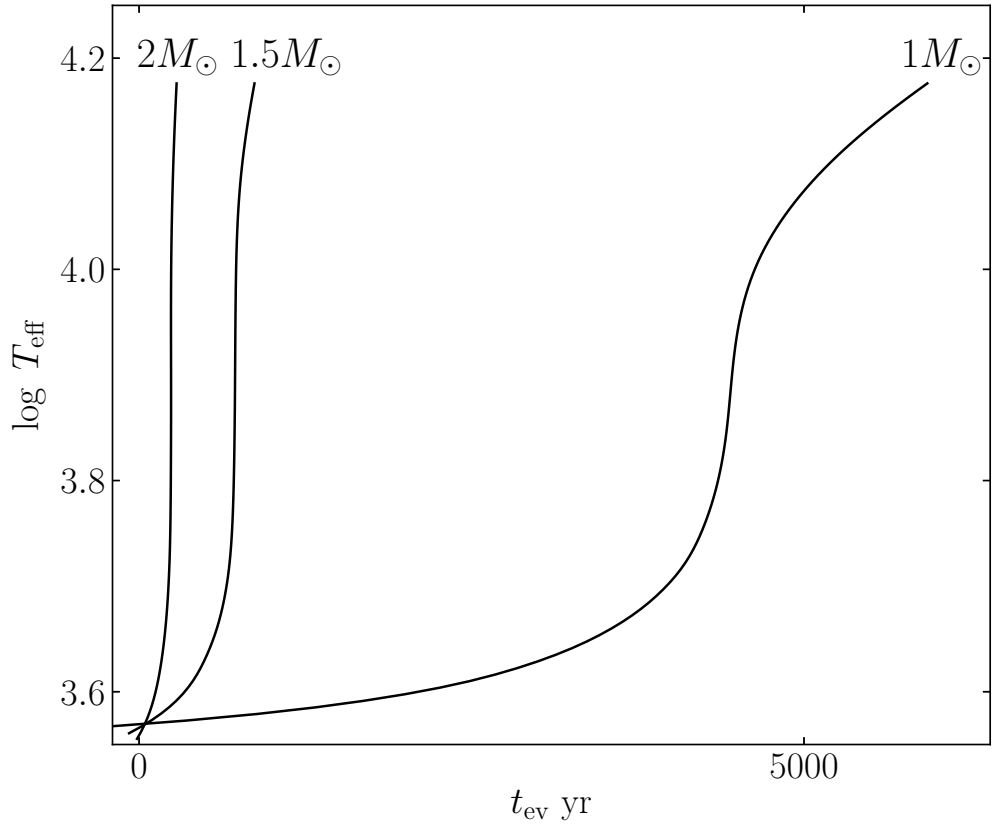}}
\caption{Effective temperature $\teff$ of post--AGB stars as a function of evolutionary time $\tev$
         for evolutionary sequences $M_0=1M_\odot$, $1.5M_\odot$ and $2M_\odot$.}
\label{fig2}
\end{figure}

The last column in Table~\ref{tabl1} gives the evolutionary time $\Delta\tev$ when
the effective temperature becomes as high as $\teff = 2\times 10^4$ K.
For the stellar initial mass increasing from $M_0=1M_\odot$ to $M_0=2M_\odot$ the
evolutionary time $\Delta\tev$ reduces by a factor of $\approx 20$.
Plots of the effective temperature $\teff$ as a function of evolutionary time $\tev$ are
shown in Fig.~\ref{fig2}.

\section*{hydrodynamic models of post--AGB stars}

The stellar radius as well as the mass of hydrogen and helium ionization zones decrease
as the post--AGB star evolves on the HRD towards the area of planetary cores.
Therefore, evolutionary decrease of the pulsation period is accompanied by decreasing
pulsation instability growth rate and diminishing role of nonlinear effects in
stellar oscillations.
At effective temperatures $\teff < 4000$\,K radial pulsations are driven in the hydrogen
ionization zone which encompasses the substantial part of the stellar envelope.
In stars with higher effective temperatures radial oscillations are driven in the helium
ionization zones.
Unfortunately, absence of strict repetition of pulsation motions does not allow us
to calculate the mechanical work $\oint PdV$ done by each mass zone of the hydrodynamic
model (here $P$ is the total pressure and $V$ is the specific volume) in order to evaluate
its contribution into excitation or suppression of pulsational instability.

As in our earlier studies devoted to nonlinear pulsations of AGB stars (Fadeyev, 2017; 2018)
the pulsation period $\Pi$ was determined using the discrete Fourier transform of the
kinetic energy of the pulsating stellar envelope.
Thus, the period estimate of each hydrodynamic model was obtained by averaging over the time
interval of the solution of the equations of hydrodynamics.
For most hydrodynamic models the solution of the Cauchy problem comprised nearly
a hundred pulsation cycles.

\begin{figure}
\centerline{\includegraphics[width=14cm]{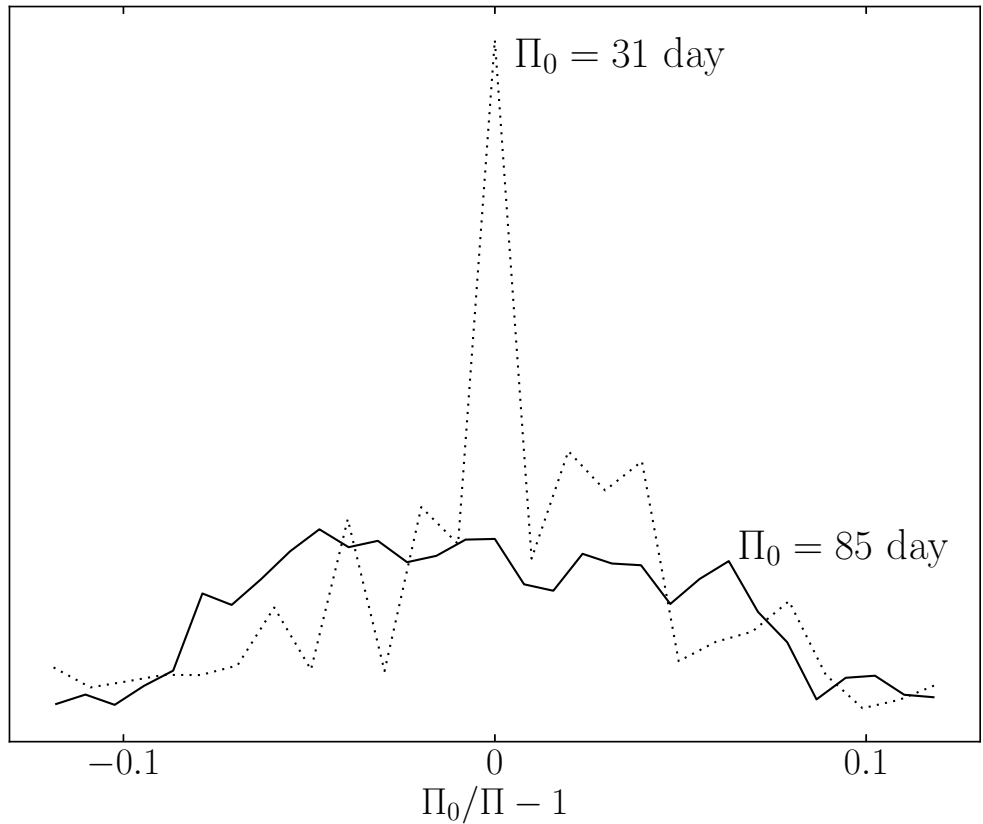}}
\caption{Normalized power spectra of the pulsation kinetic energy $\ek$ for hydrodynamic models
         with initial mass $M_0=1.5M_\odot$ at effective temperatures $\teff=4800$\,K
         (solid line) and $\teff=6500$\,K (dotted line).
         The mean period of the fundamental mode $\Pi_0$ is given near the plot.}
\label{fig3}
\end{figure}

A reduction of the role of nonlinear effects during evolution of the post--AGB star is
illustrated in Fig.~\ref{fig3} where the plots of normalized power spectrum of the pulsation
kinetic energy are shown for two hydrodynamic models of the evolutionary sequence $M_0=1.5M_\odot$.
At the stellar effective temperature $\teff=4800$\,K the growth rate of kinetic energy is
$\eta = \Pi d\ln\ek/dt = 1.3$ and the mean fundamental mode period is $\Pi_0=85$ day.
As can be seen in Fig.~\ref{fig3}, cycle--to--cycle variations of the period due to nonlinear effects
range within $\approx 20\%$ of the mean period.
In the star with effective temperature $\teff=6500$\,K the growth rate decreases up to $\eta=0.04$
whereas cycle--to--cycle variations of the period around its mean value $\Pi_0=31$ reduce
to a few per cent.

\begin{figure}
\centerline{\includegraphics[width=14cm]{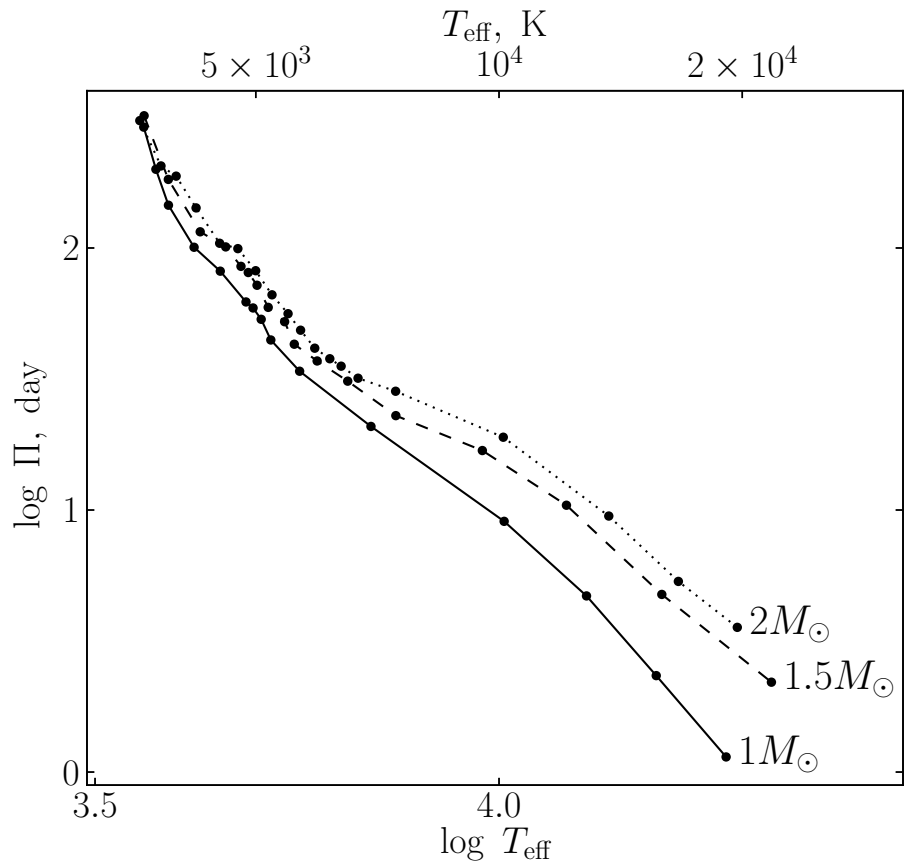}}
\caption{The period of radial pulsations $\Pi$ as a function of effective temperature $\teff$
         for evolutionary sequnces with initial masses $M_0=1M_\odot$ (solid line),
         $1.5M_\odot$ (dashed line) and $2M_\odot$ (dorred line).
         Filled circles represent mean pulsation periods of hydrodynamic models.}
\label{fig4}
\end{figure}

The plots of the mean radial pulsation period as a function of effective temperature
for evolutionary sequences $M_0=1M_\odot$, $1.5M_\odot$ and $2M_\odot$ are shown in
Fig.~\ref{fig4}.
At the beginning of the post--AGB stage the pulsation period is $\Pi\approx 300$ in
all evolutionary sequences.
While the effective temperature increases up to $\teff=2\times 10^4$\,K
the period decreases by more than two orders of magnitude.
For more convenient comparison of theoretical dependences with observations
the star age $\tev$ and the pulsation period $\Pi$ are listed for each evolutionary sequence
in Table~\ref{tabl2} as a function of the effective temperature.
The presented tabular data were obtained by nonlinear interpolation of the results of
evolutionary and hydrodynamic computations.

In the last column of Table~\ref{tabl2} we give typical amplitude of the bolometric light
$\Delta M_\mathrm{bol}$.
In stars with effective temperature $\teff < 6000$\,K modulation of radiative flux takes place in
the hydrogen ionization zone, so that the amplitude of bolometric light is $\Delta M_\mathrm{bol}\approx 1$
mag.
At effective temperatures $\teff\approx 8\times 10^3$\,K the principal role in modulation
of the radiative flux belongs to helium ionization zones and the light amplitude reduces to 
$\Delta M_\mathrm{bol}\approx 0.1$.
Further evolutionary  increase of effective temperature is accompanied by decrease of the mass
of the pulsating envelope so that the bolometric light amplitude becomes less than 0.01 mag
for $\teff > 10^4$\,K.

\section*{conclusions}

In this paper we presented results of stellar evolution and nonlinear stellar pulsation
calculations that describe the evolutionary change of the pulsation period of post--AGB stars.
Self--consistent solution of the equations of radiation hydrodynamics and time--dependent convection
allowed us to compute sequences of hydrodynamic models for effective temperatures ranging from
3600\,K to $2\times 10^4$\,K.
Nearly twentyfold change of the evolution rate within the initial mass interval
$1M_\odot\le M_0\le 2M_\odot$ allows us to conclude that most of observed post--AGB stars
originate from stars with intial mass $M_0\approx 1M_\odot$.
For more detailed comparison of the theory with observations the existing grids of
evolutionary and hydrodynamic models of post--AGB stars should be extended due to
different values of initial metal abundances $Z_0$ and mass loss rates.

Dependences given in Fig.~\ref{fig4} and in Table~\ref{tabl2} can be used for determination
of the evolutionary status of the observed pulsating variable which is suspected to be the
post--AGB star.
For example, let us consider the pulsating variable AI~CMi.
According to the General Catalogue of Variable Stars (Samus' et al., 2017)
AI~CMi is the semiregular pulsating variable of the spectral type G5Iab with period $\Pi\approx 230$ day.
Assuming that the mean effective temperatures corresponding to the spectral type of AI~CMi is
$4500\,\mathrm{K} \lesssim\teff\lesssim 5000\,\mathrm{K}$
we find from Table~\ref{tabl2} that the upper limit of the pulsation period of the post--AGB star is
nearly two times less, that is $\Pi\approx 100$ day.
It should be noted that the main obstacle in determination of the evolutionary status of the
suspected post--AGB star is uncertainty in observational estimate of the period.
Therefore, contradiction between observational estimate of the period and evolutionary status
of AI~CMi as the post--AGB star might be avoided due to smaller observational estimates of $\Pi$.

\newpage
\section*{references}

\begin{enumerate}

\item T. Aikawa, Astrophys. Space Sci. \textbf{112}, 125 (1985a).

\item T. Aikawa, Astrophys. Space Sci. \textbf{116}, 401 (1985b).

\item Т. Aikawa, Astrophys. J. \textbf{374}, 700 (1991).

\item T. Aikawa, MNRAS \textbf{262}, 893 (1993).

\item V.P. Arkhipova, V.F. Esipov, N.P. Ikonnikova, G.V. Komissarova, A.M. Tatarnikov, and B.F. Yudin,
      Astron. Lett. \textbf{35}, 764 (2009).

\item V.P. Arkhipova, N.P. Ikonnikova, and G.V. Komissarova, Astron. Lett. \textbf{36}, 269 (2010).

\item V.P. Arkhipova, O.G. Taranova, N.P. Ikonnikova, V.F. Esipov, G.V. Komissarova, V.I. Shenavrin,
      and M.A. Burlak, Astron. Lett. \textbf{42}, 756 (2016).

\item V.P. Arkhipova, N.P. Ikonnikova, V.F. Esipov, and G.V. Komissarova, Astron. Lett. \textbf{43}, 416 (2017).

\item T. Bl\"ocker, Astron. Astrophys. \textbf{299}, 755 (1995).

\item Yu.A. Fadeyev, Astrophys Space Sci. \textbf{86}, 143 (1982).

\item Yu.A. Fadeyev, Astrophys Space Sci. \textbf{100}, 329 (1984).

\item Yu.A. Fadeyev and A.V. Tutukov, MNRAS \textbf{195}, 811 (1981).

\item Yu.A. Fadeyev, Astron. Lett. 39, 306 (2013).

\item Yu.A. Fadeyev, MNRAS \textbf{449}, 1011 (2015).

\item Yu.A. Fadeyev, Astron. Lett. 43, 602 (2017).

\item Yu.A. Fadeyev, Astron. Lett. 44, 546 (2018).

\item A.B. Fokin, A. L\'ebre, H. Le Coroller, and D. Gillet, 2001, Astron. Astrophys. \textbf{378}, 546 (2001).

\item A. Gautschy, MNRAS \textbf{265}, 340 (1993).

\item N. Grevesse and A.J. Sauval, Space Sci. Rev. \textbf{85}, 161 (1998).

\item B.J. Hrivnak, S. Kwok, and K.M. Volk, Astrophys. J. \textbf{346}, 265 (1989).

\item B.J. Hrivnak, S. Kwok, and T.R. Geballe, Astrophys. J. \textbf{420}, 783 (1994).

\item B.J. Hrivnak, W. Lu, and K.A. Nault,  Astron. J. \textbf{149}, 184 (2015).

\item B.J. Hrivnak, G. Van de Steene, H. Van Winckel, W. Lu, and J. Sperauskas, Astron. J. \textbf{156}, 300 (2018).

\item N.P. Ikonnikova, O.G. Taranova, V.P. Arkhipova, G.V. Komissarova, V.I. Shenavrin,
      V.F. Esipov, M.A. Burlak, and V.G. Metlov, Astron. Lett. \textbf{44}, 457 (2018).

\item R. Kuhfu\ss, Astron. Astrophys. \textbf{160}, 116 (1986).

\item M.M. Miller Bertolami, Astron. Astrophys. \textbf{588}, A25 (2016).

\item B. Paczy\'nski, Acta Astron. \textbf{21}, 417 (1971).

\item B. Paxton, J. Schwab,  E.B. Bauer, L. Bildsten, S. Blinnikov, P. Duffell,
      R. Farmer,  J.A. Goldberg, et al., Astropys. J. Suppl. Ser. \textbf{234}, 34 (2018).

\item N.N. Samus', E.V. Kazarovets, O.V. Durlevich, N.N. Kireeva, and E.N. Pastukhova,
      Astron. Rep. \textbf{61}, 80 (2017)].

\item D. Sch\"onberner, Astron. Astrophys. \textbf{79}, 108 (1979).

\item D. Sch\"onberner, Astron. Astrophys. \textbf{103}, 119 (1981).

\item I.S. Shklovsky, Astronomicheskii Zhurnal \textbf{33}, 315 (1956).

\item Y. Tuchman, N. Sack, and Z. Barkat, Astrophys. J. \textbf{219}, 183 (1978).

\item E. Vassiliadis and P.R. Wood,  Astrophys. J. Suppl. Ser. \textbf{92}, 125 (1994).

\item K.M. Volk and S. Kwok,  Astrophys. J. \textbf{342}, 345 (1989).

\item A. Weiss and J.W. Ferguson, Astron. Astrophys. \textbf{508}, 1343 (2009).

\item P.R. Wood and D.J. Faulkner, Astrophys. J. \textbf{307}, 659 (1986).

\item J. Zalewski, Acta Astron. \textbf{35}, 51 (1985).

\item J. Zalewski, Acta Astron. \textbf{43}, 431 (1993).

\end{enumerate}

\newpage
\begin{table}
\caption{Properties of evolutionary models at the onset of the post--AGB stage}
\label{tabl1}
\begin{center}
 \begin{tabular}{crccccr}
  \hline
  $M_0/M_\odot$ & $t_*, 10^9$ yr & $M_*/M_\odot$ & $M_{\mathrm{CO},*}/M_\odot$ & $L_*/L_\odot$ & $T_{\mathrm{eff},*}$, K & $\Delta\tev$, yr \\
  \hline
1.0 & 12.607 &  0.542 &  0.493 &   3322 &   3707 &   7485 \\
1.5 &  3.040 &  0.590 &  0.550 &   5971 &   3678 &   1077 \\
2.0 &  1.352 &  0.615 &  0.581 &   8169 &   3620 &    342 \\
  \hline          
 \end{tabular}
\end{center}
\end{table}
\clearpage

\newpage
\begin{table}
\caption{The evolutionary time $\tev$ and the period of radial oscillations $\Pi$ of post--AGB stars}
\label{tabl2}
\begin{center}
 \begin{tabular}{r|rr|rr|rr|l}
  \hline
  $\teff$, K & \multicolumn{2}{c|}{$1M_\odot$} & \multicolumn{2}{c|}{$1.5M_\odot$} & \multicolumn{2}{c|}{$2M_\odot$} & $\Delta M_\mathrm{bol}$\\
             & $\tev$, yr & $\Pi$, day & $\tev$, yr & $\Pi$, day & $\tev$, yr & $\Pi$, day & \\
  \hline
  3600  & -1642  &  320.9  &  -146  &  346.4  &   -11  &  298.1  & 1.0  \\
  3800  &   953  &  180.8  &   171  &  207.1  &    77  &  207.8  &      \\
  4000  &  2191  &  123.1  &   357  &  149.5  &   128  &  185.6  &      \\
  4200  &  2866  &  100.1  &   466  &  121.5  &   161  &  145.7  &      \\
  4500  &  3456  &   82.7  &   567  &  100.7  &   192  &  104.7  &      \\
  5000  &  3957  &   57.5  &   656  &   73.4  &   220  &   81.7  &      \\
  6000  &  4307  &   29.8  &   707  &   36.7  &   237  &   39.9  & 0.5  \\
  8000  &  4461  &   15.2  &   722  &   20.9  &   240  &   26.6  & 0.1  \\
 10000  &  4630  &    9.4  &   729  &   15.7  &   241  &   19.4  & 0.01 \\
 12000  &  5033  &    5.8  &   759  &   10.7  &   252  &   13.3  &      \\
 15000  &  5927  &    2.7  &   868  &    5.7  &   282  &    7.3  &      \\
  \hline          
 \end{tabular}
\end{center}
\end{table}
\clearpage

\newpage
\clearpage

\newpage
\clearpage

\newpage
\clearpage

\newpage
\clearpage

\end{document}